# Distance learning as innovation technology of school geographical education

Myroslav J. Syvyi[1][0000-0002-3150-4848], Ordenbek B. Mazbayev[2][0000-0002-8769-1240],
Olga M. Varakuta[1][0000-0001-6705-5485], Natalia B. Panteleeva[3][0000-0001-6787-2266] and
Olga V. Bondarenko[3][0000-0003-2356-2674]

[1] Ternopil Volodymyr Hnatiuk National Pedagogical University,
2 Maxyma Kryvonosa Str., Ternopil, 46027, Ukraine
`syvyjm@ukr.net, ovarakuta@ukr.net`
[2] L. N. Gumilyov Eurasian National University,
2 Satpayev Str., Nur-Sultan, 010008, Republic of Kazakhstan
`ordenbek@mail.ru`
[3] Kryvyi Rih State Pedagogical University, 54 Gagarin Ave., Kryvyi Rih, 50086, Ukraine
`panteleeva4y@gmail.com, bondarenko.olga@kdpu.edu.ua`

**Abstract.** The article substantiates the necessity of using innovative technologies in the process of studying and teaching geographical disciplines at secondary schools. Particular attention is paid to distance learning as a pedagogical innovation, its theoretical aspects and the ways of its introduction into the educational process. The relevance of using distance learning at the New Ukrainian School is proved. Its advantages and disadvantages are revealed. The examples of some forms of distance learning that will contribute to geographical competence development according to European requirements are provided. The article particularly focuses on the Massive Open Online Courses, modern websites, virtual portals of individual teachers, LearningApps.org portal, and Moodle.

## 1  Introduction

### 1.1  The problem statement

Today's young people live in a media environment where the use of computers, Internet resources and mobile devices is part of their daily lives, they are, in the words of Alexander Kuleshov, "digitally born" [18], and this cannot be ignored. Students seemed ready for distance learning, but forced distance learning during quarantine was a challenge for all participants in the learning process: students, teachers and parents. It was very difficult to organize quality education using digital technologies, motivate students, give advice on overcoming technical problems. Global quarantine has made its unexpected adjustments and forced everyone to urgently learn digital tools and new pedagogical approaches and methods [11].

The urgency of the problem of our study can be presented in the form of a contradiction between the social requirements for the proper organization of distance learning and the state of readiness for it as individual participants in the educational process and the domestic education system as a whole.

### 1.2　Theoretical background

Distance learning has become the characteristic innovation of the last decade, which is sometimes understood by the scientists as a way of learning process implementation which involves using modern telecommunication and innovative technologies that make studying in distance without personal teacher-student contacts possible [1]; organizational form of educational process in the specially created educational environment using modern information and communication technologies [2].

Not denying the definitions, given by the scientists, taking into account the specific character of geography, let's understand distance learning as the individual process of gaining knowledge, abilities, skills and the ways of the personal cognitive activity, occurring mainly at the mediated interaction of the participants of the training process, being distant from one another, in the specialized environment, functioning on the basis of modern psychological-pedagogical and information-pedagogical technologies, in the presented publication.

The different aspects of distance learning were revealed by Aleksandr A. Andreev [1], Borje Holmberg [13], Mihail P. Karpenko [15], Desmond Keegan [17], Michael Moore [31], Vasilii I. Soldatkin [1], Charles Wedemeyer [50], and others. Pedagogical works which raise issues of theory and practice of distance learning cover the following topics of scientific research: scientific support for distance learning, problems and areas of research (Valerii Yu. Bykov [6; 7], Volodymyr M. Kukharenko [24], Yukhym I. Mashbyts [28], Natalya V. Rashevska [41], Serhiy O. Semerikov [30; 33], Andrii M. Striuk [46], Bohdan I. Shunevych [43], Nina H. Syrotenko [5], Yurii V. Tryus [48], Myroslav I. Zhaldak [53]), organizational and pedagogical conditions of distance learning, approaches to their realization (Hennadiy M. Kravtsov [23], Svitlana V. Shokaliuk [42], Vasyl V. Yahupov [51]); development and use of multimedia and computers for teaching different subjects, for instance, foreign languages (Svitlana M. Amelina [47], Olha V. Chorna [9], Vita A. Hamaniuk [16], Iryna S. Mintii [49]) and music (Liudmyla H. Havrilova [12]); marketing of distance learning (Larysa M. Petrenko [36]).

### 1.3　The objective of the article

The objective of the article is to reveal the meaning of distance learning, to characterize its specific forms and to demonstrate the features of their application in school geographic education.

## 2 Results and discussion

A retrospective analysis of scientific research suggests that the origins of distance learning as a means of education date back to the 19th century, when the University of London offered "correspondence training". The educational institution, in parallel with full-time education, introduced such forms of education that were carried out at a distance from the educational institution. The term "distance learning" was officially recognized in 1982 [17]. Distance education began to become more widespread in the 1990s in countries such as the United Kingdom, the United States and France. In Ukraine, such education began to be introduced and developed only at the end of the 20th century and mainly in the higher education system.

Distance learning has revealed itself as the most efficient system of the 21st century educational process in different educational institutions. The main objective of distance learning is to develop learners' creative and intellectual skills by means of the open and free use of all educational resources and programs.

As noted above, distance learning gains more and more popularity in our state nowadays, due to the development of the information technologies, the change of the attitude to the traditional education and also due to the announcement of the quarantine, connected with the pandemic of COVID-19. The task arose before the teaching staff, which required the urgent solvation on conditions of the quarantine: the organization of distance learning.

A little bit before, distance learning remained extremely urgent, due to the situation in the AR of the Crimea and also in the Donetsk and Lugansk special districts. Young residents of the temporarily occupied territories had not had the access to Ukrainian education for a long time and, consequently, had not had the opportunity to go to Ukrainian higher educational institutions. Therefore, in September 2016, the E-school educational portal was created where students can choose one of the five basic Ukrainian schools in Donetsk region and gain free distance education. In addition, the MES website has a list of schools from all over Ukraine where distance education can be obtained.

There were three schools in general in our state, concentrated only at distance learning, before the announcement of the quarantine: Optima Education Center (Kyiv); Distance school ANGSTREM (Kharkiv); Center for Distance Education Source (Kyiv). These schools require tuition fee. Despite functioning only few years, they are successful at implementing world distance education standards [19].

Let's characterize a distance learning geography. The basic elements of the distance learning system are the sets of educational and methodological materials from different disciplines (geography included), presented in the following forms: distance courses, interactive learning resources, set of online lessons and teaching fragments, virtual learning environments, electronic libraries, electronic circuit simulators, electronic periodicals, electronic systems for monitoring and assessing learning results, computer simulation tools, computer demonstrations.

The development of such sets is supposed to be done on the basis of the content of the State Educational Standards of each subject, and the volume and content of the sets should be sufficient or excessive for educational process, considering the students who have different basic education, different learning styles and skills.

An analysis of the experience of using distance learning in the study of geography made it possible to single out the following forms of classes that are most often used:

— chat sessions – training sessions that are conducted using chat technologies. Chat sessions involve synchronous interaction what means that all participants have instant access to the chat. Many distant educational establishments operate chat-schools where teacher-student interaction is organized by means of chat rooms [38];
— web-classes – distant lessons, conferences, seminars, business games, and other forms of telecommunication and Internet training;
— video lessons are an integral part of distance learning. Digital files can be stored on an individual electronic device or on a web server. Typically, a record of teaching process is broadcast on the screen. It is considered effective for distance learning to use dynamic video aids: movie clips, animations, spreadsheets. The advantage of this form of material presentation is the opportunity for a student to regulate the course of the lesson individually, to review previous stages and difficult moments;
— web-forums – the form of dealing with a particular topic or issue by means of posts that remain on one of the sites with the relevant program installed on it. Web forums are distinguished from chat sessions by longer (multi-day) work and the asynchronous nature of teacher-student interaction;
— distant conference as a class can take several forms, therefore, the following types are distinguished: video conference, audio conference computer conference, and teleconference. Teleconference is usually carried out on the basis of mailing lists using e-mail [19];
— individual work (investigative, creative) based on a certain plan, schedule or scenario;
— individual projects;
— trainings, master-classes, workshops;
— assessment (tests, keys to tests);
— consultations.

The specific educational need of a modern subject teacher is to create a virtual portal and integrate it into the single information space. This Interactive Geographer's Office was created on Iryna V. Makhanko's site [27]. The distance course "The Last Terra Incognita of Our Planet" is designed for students' individual studying, comprehension, knowledge enlargement and self-evaluation of all geographical courses. For example, in the course "Geography of continents and oceans" in the part "South America is a mysterious lost world", the motto of which is: "A person cannot truly improve themselves unless they help others to develop", C. Dickens, the course developer singled out different sections with hyperlinks to the information and tasks, among which are: "Case for creative students", "Co-operative tables to the topic", "Teacher's Web consultation", "Collective opinion: voting", "Collective diary of the explorer", "Advanced tasks, materials for self-study", "Topic assessment: online testing".

The author of the site has offered other rubrics to study the topic "Empire of Cold Incognita": "Program and explanations for the course", "Interactive map", "Nature of Antarctica", "Cartographic simulator", "Video excursion to the course", "Teacher's Web consultation", "Geographical workshop", "Scaffolding schemes for classes", "Test yourself", "Questions for self-control", "Topics of research works", "Useful links", "Course test – online testing".

The course algorithm is easy to use. The content of the rubrics is accessible, scientific, systematic, visually enriched.

The use of elements of automated learning, Internet portals enables teachers to increase the students' cognitive interest in self-study of geographical content.

Geography teachers using innovative teaching methods understand the simplicity and effectiveness of modern educational tools such as LearningApps.org [25].

Registration to this portal is required for both teachers and students. To work on the tasks at individual pace and to have this process controlled, the users may apply a special program. To use it, they need to connect their computers to a local Internet network.

On the portal home page, there is a video with brief instructions of how to work with the portal; there are several windows with the examples of exercises for different subjects and of different levels; the following sections are attached: "What is LearningApps.org", "Show help". Also, on this page there are sections with the access to other windows with additional information: "About LearningApps.org", "About us", "Agreements / Terms", the icon for registration, review of exercises, designing exercises, a search window – "Exercise view".

The LearningApps.org has the following characteristics:

- the option to create individually all kinds of puzzles, crosswords, matching tasks, games of cartographic format, classifications, simple ordering, quizzes, etc.;
- easy and comfortable use, interface is simple for navigation;
- comfortable language choice, Ukrainian included;
- the choice of the task level: from the pre-school to postgraduate levels;
- the opportunity to do the tasks presented at the portal for enhancing knowledge;
- transferring the content of the exercises to registered users, including teachers and students, either simultaneously to all students' workplaces, or selectively to some of them. It gives the opportunity to consult students during the creation of certain exercises, to check the level of the acquired knowledge and skills, to view the results of the done exercises;
- the option to register under both teacher's and student's profile with the opportunity of organizing classes with the aligned conversations [45].

LearningApps.org is a service to support learning and teaching processes through small interactive modules. These modules can be used either directly as training resources or for independent work.

You can create your account under the title "Create an account" providing the required data: login (phone number), e-mail, password. Registered users can design their own exercises. To do this, select the section "Creating exercises" on the home page. The suggested rubrics are: "Matching", "Classification", "Ordering", "Long text answer", "Parts of a picture", "Multiple choice", "Fill in the gaps", "Exercise collection", "Audio- and video collection", "First million", "Puzzle", "Crossword", "Find the word", "Where is it?", "Guess the word", "Jumps", "Pairs".

The further preview or completion of the creating exercises online is possible. It is advisable to use them at the stage of testing, generalization and systematization or application of the acquired knowledge and skills, or at the stage of their direct formation and acquisition. It is also possible to suggest designing similar exercises for students as homework.

In the modern world, blogs and other types of websites are used to communicate and interact with the participants of the educational process. Such browsers as Mozilla Firefox, Opera, Google Chrome, Safari make it possible [37].

Blog (English origin) is a modern web-site with the continuously added posts as the main content that contain texts, images, presentations, films etc. For authors, blogs provide an opportunity to create materials, share important information and experience, conduct online testing, organize team work, etc. For users, blogs are examples of creative tasks, participation in projects, competitions, means for posting photos, notes, creative works and for communication to discuss common issues and, what is more, it is an opportunity for self-education [39].

Blogs and websites play an important role in the educational process for teaching and studying geographical disciplines serving as a source of information published on the site and used to organize distance learning; monitoring; work and task discussion.

The example of Geography distance learning can be Olena Chuiko's (Geography teacher) blog (https://geovsviti.blogspot.com/). Its content consist of the following rubrics: the main (geographical events, nationally patriotic education, recommendations); author's portfolio; photos (made by the author concerning the local area and learning activities), methodological materials (textbooks, manuals, geographical problems); visual aids (posters, maps to various geography courses and topics), presentations (on various geographical topics), lessons (the information is presented using various visual, practical, multimedia materials), practical techniques (activities like "Jigsaw", "STEM-lesson", "Flash-cards", etc.); contests, videos (on the topics "How to create a blog", "Travelers", "Ukraine", "Countries of the world", "Continents", "Oceans", "Famous buildings"). The author demonstrates the ways of implementing the experience into the practice.

"Geography for the curious" is a personal blog of Geography and Economics teacher of Pereyaslav-Khmelnitsky gymnasium Olga Chemeris [8]. The analysis of the section "About me" proves the author to be an enthusiastic teacher using modern technologies and innovative training tools.

A number of Internet sites, including "Theory of Geography" (https://sites.google.com/site/teoriageografiie/), "Geographic. Geographical Portal" (http://geografica.net.ua), Popular Geography (http://www.geosite.com.ru/), "World of Mysteries and Wonders" (http://chudesa-sveta.narod.ru /) are advised to be used for Geography distance learning.

The world tendency is that YouTube channel is rapidly gaining popularity among users and now it is the third most popular web-site after Google and Facebook [9; 37]. Plenty of interesting and useful geographical presentations are available on the channel "Geography presentations". The playlists can demonstrate online presentations on any topic including "Environment", "Geography of Ukraine", "Industry", "Interesting facts", "World geography", "Bodies of Water", "Cartography", "Continents and Oceans" etc. The channel represents itself as a video resource for those interested in the present situation and provides students with the opportunities to do learning activities independently in order to obtain additional sources of information during distance learning.

One of the forms of an educational material set for Geography distance learning is Massive Open Online Courses (MOOC) – (substantial, massive, accessible, public, open distant online courses) – they are online courses with large-scale interactive and information participation and open Internet access [35].

The point of such courses is to provide the opportunity to use an interactive online forum which, in turn, helps to organize a union of teachers and students, as a supplement to traditional geography teaching methods such as: working with atlas maps; analyzing statistical and text tables, charts, graphs; work with a textbook, explanation and various types of conversations.

Currently, mass online courses are one of the most widely used elements of distance learning that is rapidly developing in the global education system. There are a number of widely popular mass online courses that attract broad audience and active users. The following courses are extremely popular: Coursera, EdX, Udacity. On the basis of Taras Shevchenko National University of Kyiv, mass open online courses – "Online University" – were developed in 2014. The interactive online educational project EdEra which continuously develops online resources and accessible educational content was launched [14].

Plenty of various free learning management systems are available today: Acollab, ATutor, Claroline, Colloquia, DodeboLMS, Dokeos, ELEDGE, Ganesha, ILIAS, LAMS, LON-CAPA, LRN, MOODLE, OLAT, OpenACS, OpenCartable, OpenLMS, SAKAI, The Manhattan Virtual Classroom. However, Google Classroom and Moodle are the most popular and massively used systems [20; 29].

Google Classroom is the online educational interactive tool, with the help of which you may create the educational environment, enriched by information, where the text editor Docs, the cloud storage Drive, Gmail and other addendums are combined (YouTube, Sheets, Slides, Forms, etc.) [3]. On conditions of the interactive online cooperation, Google Classroom gives you the opportunity to realize the efficient interaction of the training subjects in the real-time mode by the creation of the task for each definite class with the hyper-inclination to the multimedia content; the editing and commenting on the state of the tasks' fulfillment by the pupils; the integration of the individual tasks into the topical modules; the publication of the announcement, question, the information digests, etc.; the control's realization over the fulfillment of the individual tasks by pupils; the setting of the fulfillment terms for each task; the commenting on the reviewed multimedia content, suggested for the tasks; the evaluation of the pupils' educational achievements; the copying of the training achievements into the Google tables for the creation of the statistical reports, the visual monitoring of training's quality. The experience of the distance learning's organization at the study of geographical disciplines has been presented in details in the previous publications [3; 4].

The Moodle platform is equally popular among the teaching staff learning environment, also called the Learning Management System (LMS), the Course Management System (CMS), the Virtual Learning Environment, or just a learning platform that provides teachers, students and administrators with a very broad set of tools for computerized learning, including distance learning. It is the most advanced and widespread system in Ukraine and in the world used for this purpose. At the moment, Moodle continues to progress at a much faster pace than its competitors. At schools where the system is used, teachers (Geography teachers included), students, and parents are all connected to it.

Moodle system enables the users to implement the following communication mechanisms: perceptional (responsible for each other's perception); interactive (responsible for arranging interaction); communicative (responsible for sharing information).

Its use as a means of distance learning has a number of advantages, namely: considerable motivational potential; confidentiality; higher levels of interactivity compared to classroom work; absence of "error fear"; opportunity for repeated material revision; modularity; dynamic access to information; accessibility; availability of a constant active help system; option of self-control; compliance with the principles of developmental learning; individualization; providing visibility and variability of information presenting.

Having analyzed the experience of the distance learning organization by the educational establishments, being promulgated on their web-sites, we tried to classify the means and the technologies of distance learning (table 1).

**Table 1.** Tools and technologies of distance learning, that gained widening on conditions of quarantine [40; 44].

| Distance learning tools | Examples |
|---|---|
| LMS | CenturyTech, ClassDojo, Google Classroom, Moodle, Schoology |
| Videoconferencing | Zoom, Skype, Teams, Hangouts Meet, Lark, Dingtalk |
| Messengers | Viber, Telegram, WhatcApp |
| Social networks | Facebook, Instagram, Twitter, Pinterest, Reddit |
| Educational projects | To the Lesson. Vseosvita, EdPro |
| Teacher's blogs and the interactive rooms | O. Chuiko (https://geovsviti.blogspot.com/, I. Makhanko (https://mo-teachers-sc4.ucoz.com/), M. Maznytsya (https://mariwamazni4ka.wixsite.com/mysite), V. Kotsybynska (https://sites.google.com/view/kotsyubinskaya/) |
| Educational sites | "Theory of Geography" (https://sites.google.com/site/teoriageografiie/), Site "Geographics. Geographical Portal" (http://geografica.net.ua), Site "Popular Geography" (http://www.geosite.com.ru/), Site "World of Mysteries and Wonders" (http://chudesa-sveta.narod.ru/) |
| Electronic libraries | Electronic Library of Ukraine, UkrLib, Open Book, Library of Originals and Translations of the World Literature Compositions, Poetics, Portal, OpenLibrary, Pensilvania University Library, etc. |
| Online training contents | Byju's, Discovery Education, Khan Academy, KitKit School, LabXchange, Mindspark, OneCourse, Quizlet (link is external) |

377377

Distance learning has both the advantages and disadvantages (table 2).

**Table 2.** Advantages and disadvantages of distance learning on conditions of quarantine [10; 21; 22; 32].

| Advantages | Disadvantages |
|---|---|
| — the ability to get knowledge anytime and anywhere;<br>— the psychological stability and comfort of training. ability to work at a pace, being convenient for a pupil or a student, and the organization of the individual approach to everybody;<br>— the possibility to control the educational process from the side of parents;<br>— the availability of educational materials;<br>— the interest of pupils and students in the use of new means and tools for learning;<br>— the convenience for the teacher;<br>— the possibility of educational process centralization. | — the access inability of pupils and students to high-quality Internet (especially in the rural area);<br>— the absence of a single platform for the organization of distance learning;<br>— the possibility of hacker invasion into the electronic database;<br>— the insufficient amount of equipment (the age or the software correspondingly) in the families, where there are pupils and students, the high cost of equipment for the realization of distance learning;<br>— the contents of distance learning may depend on the technological characteristics of the equipment used;<br>— the absence of practical abilities and skills of application of both pupils and teachers;<br>— the predominance of the external motivation of learning and the low readiness level of separate pupils and teachers for the work in a new environment;<br>— the absence of some pupils' skills of the independent mastering of educational material;<br>— the insufficient control over the pupils' mastering of the acquired knowledge. a great amount of time for the prepared works' checking by the teacher. the problem of estimation;<br>— copyright problems in the use of educational materials;<br>— distance education is not suitable for the development of the sociability;<br>— the problem of the pupil's identification. |

To overcome the problems, appeared at distance learning, we may offer the following ways of their solution [26; 32; 34; 52]:

— to provide an equal access to the high-quality Internet connection and to the technical means for the education getters and teachers;
— to elaborate the single standard for distance learning;
— the educational establishments should operate on one platform;
— to improve the qualification of teachers, dealing with the use of ICT-technologies;
— to increase the education getters' motivation to training;
— to renovate the modern educational programs;
— to elaborate the single standards for the estimation and control of the training achievements of the education getters;
— to renovate the material and technical base of educational establishments;
— to reduce the educational loading for the education getters at the expense of application of the new modern methods of training.

## 3  Conclusions

1. The introduction of innovative methods in natural sciences teaching allows to change radically the approaches to the educational process. The broad introduction of innovations is facilitated by the rapid development of information and communication technologies, which creates new educational opportunities. The use of computer technologies enhances the educational process, provides new ways of acquiring information, provides testing for new ideas and projects.
2. Distance learning is the individual process of gaining knowledge, abilities, skills and the ways of the personal cognitive activity, occurring mainly at the mediated interaction of the participants of the training process, being distant from one another, in the specialized environment, functioning on the basis of modern psychological-pedagogical and information-communication technologies. Distance learning is actively used as an innovative technique for the geographical competence.
3. On the basis of the content of the State Educational Standard for each subject (geography included), sets of educational and methodical materials are being developed and presented in the form of distant courses, interactive training resources, online lessons, electronic simulators, electronic periodicals, electronic systems of monitoring and assessment of educational results, etc. Distance learning via computer telecommunications has the following forms: chat, web-lessons, video-lessons, web -forums, distant conferences, individual project works; trainings, etc.
4. The most widely used means of distance learning nowadays are considered to be Massive Open Online Courses (Coursera, EdX, Udacity, EdEra, etc.), virtual web portals (LearningApps.org., Teachers of Geography Online), websites (Blogs, Google, Facebook, YouTube), Learning Management Systems (Acollab, ATutor, Claroline, Colloquia, Dokeos, ELEDGE, Ganesha, ILIAS, LAMS, LON-CAPA, LRN, Moodle, OLAT, OpenACS, OpenCartable, OpenLMS, SAKAI, The Manhattan Virtual Classroom) etc. The most advanced and widespread both in Ukraine and around the world is viewed to be the Moodle system, which enables users to implement all the basic mechanisms of communication.
5. Distance learning is characterized by a number of advantages: the chance to study at convenient time, anywhere and at individual pace; the opportunity to use information from multiple sources, media files, teachers' comments, article links, etc.; sparing time for extracurricular activities; encouragement to acquire skills to use modern information and communication technologies.
6. The problems with the wide application of distance learning in Ukrainian geographical educational process are caused mainly by insufficient school supply of modern technological aids, Internet, as well as methodological, psychological and pedagogical recommendations concerning distant learning, high requirements for a "virtual" teacher.
7. We see the perspectives of the further scientific searching in the elaboration of the model and the methods of the future Geography teachers' preparation to distance learning.

## References

1. Andreev, A.A., Soldatkin, V.I.: Dystantsyonnoe obuchenye: sushchnost, tekhnolohyia, orhanyzatsyia (Distance Learning: Essence, Technology, Organization). MESY, Moskva (2009)
2. Andreev, A.A.: Didakticheskie osnovy distantcionnogo obucheniia v vysshikh uchebnykh zavedeniiakh (Didactic bases of distance learning in higher educational institutions). Dissertation, Moscow State University (1999)
3. Bondarenko, O.V., Mantulenko, S.V., Pikilnyak, A.V.: Google Classroom as a Tool of Support of Blended Learning for Geography Students. In: Kiv, A.E., Soloviev, V.N. (eds.) Proceedings of the 1st International Workshop on Augmented Reality in Education (AREdu 2018), Kryvyi Rih, Ukraine, October 2, 2018. CEUR Workshop Proceedings **2257**, 182–191. http://ceur-ws.org/Vol-2257/paper17.pdf (2018). Accessed 30 Nov 2018
4. Bondarenko, O.V., Pakhomova, O.V., Lewoniewski, W.: The didactic potential of virtual information educational environment as a tool of geography students training. In: Kiv, A.E., Shyshkina, M.P. (eds.) Proceedings of the 2nd International Workshop on Augmented Reality in Education (AREdu 2019), Kryvyi Rih, Ukraine, March 22, 2019. CEUR Workshop Proceedings **2547**, 13–23. http://ceur-ws.org/Vol-2547/paper01.pdf (2020). Accessed 10 Feb 2020
5. Bykov, V. Yu., Kukharenko, V.M., Syrotenko, N.H., Rybalko, O.V., Bohachkov, Yu.M.: Tekhnolohiia stvorennia dystantsiinoho kursu (Technology of creation of a distance course). Milenium, Kyiv (2008)
6. Bykov, V., Dovgiallo, A., Kommers, P.A.M.: Theoretical backgrounds of educational and training technology. International Journal of Continuing Engineering Education and Life-Long Learning **11**(4-6), 412–441 (2001)
7. Bykov, V., Gurzhiy, A., Kozlakova, G.: Development of computer education in Ukrainian higher technical schools. In: IFIP Transactions A: Computer Science and Technology (A-52), pp. 678-681 (1994)
8. Chemerys, O.: Heohrafiia dlia dopytlyvykh (Geography for the curious). http://chemerusjlga.blogspot.com/p/blog-page_10.html. Accessed 27 Feb 2020
9. Chorna, O.V., Hamaniuk, V.A., Uchitel, A.D.: Use of YouTube on lessons of practical course of German language as the first and second language at the pedagogical university. In: Kiv, A.E., Soloviev, V.N. (eds.) Proceedings of the 6[th] Workshop on Cloud Technologies in Education (CTE 2018), Kryvyi Rih, Ukraine, December 21, 2018. CEUR Workshop Proceedings **2433**, 294–307. http://ceur-ws.org/Vol-2433/paper19.pdf (2019). Accessed 25 Oct 2019
10. Danylo, L.: Dystantsiina osvita ta umovy yii vprovadzhennia (Distance learning and the conditions of its implementation). Visnyk Luhanskoho natsionalnoho universytetu **4**, 59-66 (2018)
11. Hamaniuk, V., Semerikov, S., Shramko, Y.: ICHTML 2020 – How learning technology wins coronavirus. In: Hamaniuk, V., Semerikov, S., Shramko, Y. (eds.) The International Conference on History, Theory and Methodology of Learning (ICHTML 2020). Kryvyi Rih, Ukraine, May 13-15, 2020. SHS Web of Conferences **75**, 00001 (2020). doi:10.1051/shsconf/20207500001
12. Havrilova, L.H., Ishutina, O.Ye., Zamorotska, V.V., Kassim, D.A.: Distance learning courses in developing future music teachers' instrumental performance competence. In: Kiv, A.E., Soloviev, V.N. (eds.) Proceedings of the 6[th] Workshop on Cloud Technologies in Education (CTE 2018), Kryvyi Rih, Ukraine, December 21, 2018. CEUR Workshop

Proceedings **2433**, 429–442. http://ceur-ws.org/Vol-2433/paper29.pdf (2019). Accessed 10 Sep 2019
13. Holmberg, B.: Guided didactic conversation in distance education. In: Sewart, D., Keegan, D., Holmberg, B. (eds.) Distance Education: International Perspectives. Routledge, London and New York (1988)
14. Interaktyvna onlain-osvita (Interactive online education). EdEra. https://www.ed-era.com (2020). Accessed 26 Feb 2020
15. Karpenko, M.P.: Sovremennaya didaktika massovogo elektronnogo obrazovaniya (Modern didactics of mass electronic education). Sovremennyye informatsionnyye tekhnologii i IT-obrazovaniye 13, 58–64 (2017). DOI:10.25559/SITITO.2017.4.560
16. Kazhan, Yu.M., Hamaniuk, V.A., Amelina, S.M., Tarasenko, R.O., Tolmachev, S.T.: The use of mobile applications and Web 2.0 interactive tools for students' German-language lexical competence improvement. In: Kiv, A.E., Shyshkina, M.P. (eds.) Proceedings of the 7[th] Workshop on Cloud Technologies in Education (CTE 2019), Kryvyi Rih, Ukraine, December 20, 2019. CEUR Workshop Proceedings **2643**, 392–415. http://ceur-ws.org/Vol-2643/paper23.pdf (2020). Accessed 20 Jul 2020
17. Keegan, D.: Foundations of distance education, 3[rd] edn. Routledge, New York (1996)
18. Konstantinov, A.: Cifrorozhdjonnye: Aleksandr Kuleshov o primetah budushhego, cifrovom intellekte i pokolenii Z – ljudjah, vospitannyh gadzhetami (Digital-born // Alexander Kuleshov on the signs of the future, digital intelligence and generation Z - people brought up by gadgets). Kot Shrjodingera 1–2(39–40). https://kot.sh/statya/3821/cifrorozhdyonnye (2018). Accessed 20 December 2018
19. Korbut, O.H.: Dystantsiine navchannia: modeli, tekhnolohii, perspektyvy (Distance Learning: models, techniques, prospects. http://confesp.fl.kpi.ua/ru/node/1123 (2013). Accessed 25 Feb 2020
20. Korobeinikova, T.I., Volkova, N.P., Kozhushko, S.P., Holub, D.O., Zinukova, N.V., Kozhushkina, T.L., Vakarchuk, S.B.: Google cloud services as a way to enhance learning and teaching at university. In: Kiv, A.E., Shyshkina, M.P. (eds.) Proceedings of the 7[th] Workshop on Cloud Technologies in Education (CTE 2019), Kryvyi Rih, Ukraine, December 20, 2019. CEUR Workshop Proceedings **2643**, 106–118. http://ceur-ws.org/Vol-2643/paper05.pdf (2020). Accessed 20 Jul 2020
21. Korolchuk, M.: Dystantsiine navchannia: za i proty. Osvita ditei (Distance learning: pros and cons. Children's education). https://learning.ua/blog/201810/dystantsiine-navchannia-za-i-proty/ (2018). Accessed 26 Feb 2020
22. Kozhevnykova, A.: Dystantsiyne navchannya yak innovatsiyna pedahohichna systema ta element kompetentnosti suchasnoho vchytelya (Distance learning as innovative pedagogical system and the element of modern teacher's competence). Imidzh suchasnoho pedahoha **4**, 14–16 (2015)
23. Kozlovsky, E.O., Kravtsov, H.M.: Multimedia virtual laboratory for physics in the distance learning. In: Semerikov, S.O., Shyshkina, M.P. (eds.) Proceedings of the 5th Workshop on Cloud Technologies in Education (CTE 2017), Kryvyi Rih, Ukraine, April 28, 2017. CEUR Workshop Proceedings **2168**, 42–53. http://ceur-ws.org/Vol-2168/paper7.pdf (2018). Accessed 21 Mar 2019
24. Kukharenko, V., Oleinik, T.: Open distance learning for teachers. CEUR Workshop Proceedings **2393**, 156-169 (2019)
25. LearningApps - interactive and multimedia learning blocks. https://learningapps.org (2020). Accessed 21 Mar 2020

26. Lyashenko, I.V.: Perspektyvy rozvytku dystantsiynoho navchannya u vyshchiy shkoli (Prospects for the Development of Distance Learning in Higher Education). Narodna Osvita 1(25). https://www.narodnaosvita.kiev.ua/?page_id=2682 (2015).Accessed 21 Mar 2020
27. Makhanko, I.V.: Interaktyvnyy kabinet heohrafa (Interactive Geographer's Office). https://mo-teachers-sc4.ucoz.com (2020). Accessed 24 Feb 2020
28. Mashbyts, Yu.I.: Teoretyko-metodolohichni zasady proektuvannia dystantsiinykh navchalnykh seredovyshch (Theoretical and methodological foundations of designing distance learning environments). In: Smulson, M.L. (ed.) Dystantsiine navchannia: psykholohichni zasady (Distance learning: psychological foundations), pp. 8–41. Imeks-LTD, Kirovohrad (2012)
29. Mintii, I.S.: Using Learning Content Management System Moodle in Kryvyi Rih State Pedagogical University educational process. In: Kiv, A.E., Shyshkina, M.P. (eds.) Proceedings of the 7th Workshop on Cloud Technologies in Education (CTE 2019), Kryvyi Rih, Ukraine, December 20, 2019. CEUR Workshop Proceedings **2643**, 293–305. http://ceur-ws.org/Vol-2643/paper17.pdf (2020). Accessed 20 Jul 2020
30. Modlo, Ye.O., Semerikov, S.O.: Xcos on Web as a promising learning tool for Bachelor's of Electromechanics modeling of technical objects. In: Semerikov, S.O., Shyshkina, M.P. (eds.) Proceedings of the 5th Workshop on Cloud Technologies in Education (CTE 2017), Kryvyi Rih, Ukraine, April 28, 2017. CEUR Workshop Proceedings **2168**, 34–41. http://ceur-ws.org/Vol-2168/paper6.pdf (2018). Accessed 21 Mar 2020
31. Moore, M.G.: Editorial: Distance Education Theory. American Journal of Distance Education **5**(3), l-6 (1991). doi:10.1080/08923649109526758
32. Myronov, Yu.B.: Perevahy ta nedoliky dystantsiynoho navchannya (Advantages and disadvantages of distance learning). https://kerivnyk.info/perevahy-ta-nedoliky-dystantsijnoho-navchannya (2020). Accessed 21 Mar 2020
33. Nechypurenko, P.P., Semerikov, S.O.: VlabEmbed – the New Plugin Moodle for the Chemistry Education. In: Ermolayev, V., Bassiliades, N., Fill, H.-G., Yakovyna, V., Mayr, H.C., Kharchenko, V., Peschanenko, V., Shyshkina, M., Nikitchenko, M., Spivakovsky, A. (eds.) 13th International Conference on ICT in Education, Research and Industrial Applications. Integration, Harmonization and Knowledge Transfer (ICTERI, 2017), Kyiv, Ukraine, 15-18 May 2017. CEUR Workshop Proceedings **1844**, 319–326. http://ceur-ws.org/Vol-1844/10000319.pdf (2017). Accessed 21 Mar 2020
34. Opanasiuk, Iu.: Dystantsiyne navchannya vnaslidok tradytsiynoyi evolyutsiyi systemy (Distant learning as a result of traditional system evolution). Vyshcha osvita Ukrayiny 1, 49–53 (2016)
35. Panchenko, L.F., Muzyka, I.O.: Analytical review of augmented reality MOOCs. In: Kiv, A.E., Shyshkina, M.P. (eds.) Proceedings of the 2nd International Workshop on Augmented Reality in Education (AREdu 2019), Kryvyi Rih, Ukraine, March 22, 2019. CEUR Workshop Proceedings **2547**, 168–180. http://ceur-ws.org/Vol-2547/paper13.pdf (2020). Accessed 10 Feb 2020
36. Petrenko, L., Kravets, S., Bazeliuk, O., Maiboroda, L., Muzyka, I.: Analysis of the current state of distance learning in the vocational education and training institutions. In: Semerikov, S., Chukharev, S., Sakhno, S., Striuk, A., Osadchyi, V., Solovieva, V., Vakaliuk, T., Nechypurenko, P., Bondarenko, O., Danylchuk, H. (eds.) The International Conference on Sustainable Futures: Environmental, Technological, Social and Economic Matters (ICSF 2020). Kryvyi Rih, Ukraine, May 20-22, 2020. E3S Web of Conferences **166**, 10010 (2020). doi:10.1051/e3sconf/202016610010
37. Pokryshen, D.A., Prokofiev, E.H., Azaryan, A.A.: Blogger and YouTube services at a distant course "Database management system Microsoft Access". In: Kiv, A.E., Soloviev,

V.N. (eds.) Proceedings of the 6th Workshop on Cloud Technologies in Education (CTE 2018), Kryvyi Rih, Ukraine, December 21, 2018. CEUR Workshop Proceedings **2433**, 516–528. http://ceur-ws.org/Vol-2433/paper35.pdf (2019). Accessed 10 Sep 2019

38. Priadko, A.O., Osadcha, K.P., Kruhlyk, V.S., Rakovych, V.A.: Development of a chatbot for informing students of the schedule. CEUR Workshop Proceedings **2546**, 128–137 (2019)
39. Prykhodko, A.M., Rezvan, O.O., Volkova, N.P., Tolmachev, S.T.: Use of Web 2.0 technology tool - educational blog - in the system of foreign language teaching. CEUR Workshop Proceedings **2433**, 256–265 (2019)
40. Radeiko, R.: Platformy ta instrumenty dlya navchannya onlayn (Online training platform and tools). http://aphd.ua/platformy-ta-instrumenty-dlia-navchannia-onlain (2020). Accessed 21 Mar 2020
41. Rashevska, N.V., Soloviev, V.N.: Augmented Reality and the Prospects for Applying Its in the Training of Future Engineers. CEUR Workshop Proceedings **2257**, 192–197 (2018)
42. Shokaliuk, S.V., Bohunenko, Ye.Yu., Lovianova, I.V., Shyshkina, M.P.: Technologies of distance learning for programming basics lessons on the principles of integrated development of key competences. CEUR Workshop Proceedings **2643**, 548–562 (2020)
43. Shunevych, B.: Ukraine Open University: Its prospects in distance education development. International Review of Research in Open and Distance Learning **2**(2), 168–181 (2002)
44. Shvadchak, N.: 35 instrumentiv dlya dystantsiynoho navchannya – dobirka NUSH (35 distance learning tools - NUSH selection). https://nus.org.ua/articles/30-instrumentv-dlya-dystantsijnogo-navchannya-dobirka-nush (2020). Accessed 21 Mar 2020
45. Strilets, S. I: Innovatsii u vyshchii pedahohichnii osviti: teoriia i praktyka (Innovations in higher pedagogical education: theory and practice). Chernihiv (2013)
46. Striuk, A.M., Rassovytska, M.V., Shokaliuk, S.V.: Using Blippar Augmented Reality Browser in the Practical Training of Mechanical Engineers. CEUR Workshop Proceedings **2104**, 412–419 (2018)
47. Tarasenko, R.O., Amelina, S.M., Kazhan, Yu.M., Bondarenko, O.V.: The use of AR elements in the study of foreign languages at the university. In: Burov, O.Yu., Kiv, A.E. (eds.) Proceedings of the 3rd International Workshop on Augmented Reality in Education (AREdu 2020), Kryvyi Rih, Ukraine, May 13, 2020, CEUR-WS.org, online (2020, in press)
48. Tryus, Yu.V., Herasymenko, I.V.: Approaches, models, methods and means of training of future IT-specialists with the use of elements of dual education. Journal of Physics: Conference Series (2020, in press)
49. Ustinova, V.O., Shokaliuk, S.V., Mintii, I.S., Pikilnyak, A.V.: Modern techniques of organizing computer support for future teachers' independent work in German language. CEUR Workshop Proceedings **2433**, 308–321 (2019)
50. Wedemeyer, C.A.: Learning at the back door: reflections on non-traditional learning in the lifespan. University of Wisconsin Press, Madison (1981)
51. Yahupov, V.V., Kyva, V.Yu., Zaselskiy, V.I.: The methodology of development of information and communication competence in teachers of the military education system applying the distance form of learning. CEUR Workshop Proceedings **2643**, 71–81 (2020)
52. Yatsenko, H.: Komunikatyvnist v systemi dystantsiinoho navchannia: faktory intensyfikatsii (Communication in the system of distant learning). Dissertation, Kyiv (2009)
53. Zhaldak, M.I., Franchuk, V.M., Franchuk, N.P.: Some applications of cloud technologies in mathematical calculations. Journal of Physics: Conference Series (2020, in press)